\documentclass[conference]{IEEEtran}
\usepackage{ifpdf}

\usepackage{cite}

\usepackage{xspace}

\ifCLASSINFOpdf
  \usepackage[pdftex]{graphicx}
  \graphicspath{{figures/}}
  \DeclareGraphicsExtensions{.pdf,.jpg,.png}
\else
\fi

\usepackage{stfloats}
\fnbelowfloat
\usepackage{url}

\newcommand{\lumi}{$\times10^{34} \rm \ cm^{-2} s^{-1}$\xspace }

\newcommand{\pT}{$p_{\rm T}$\xspace}

\begin{document}
\title{The Phase-1 Upgrade of the \\ATLAS Level-1 Endcap Muon Trigger}

\author{\IEEEauthorblockN{Shunichi Akatsuka\IEEEauthorrefmark{1}, on behalf of ATLAS Collaboration}
\IEEEauthorblockA{\IEEEauthorrefmark{1}Kyoto University Graduate School of Science, Kyoto, Japan }}


\maketitle

\begin{abstract}
For Run~3 (from 2021), the LHC will undergo a significant increase in instantaneous luminosity to 1.5 times its current value 
which will lead to larger collected statistics and an enhanced sensitivity to new physics.
The Phase-1 upgrade of the ATLAS Level-1 endcap muon trigger is essential to keep the physics acceptance at Run~3.
A new trigger logic to take coincidence with detectors inside and outside the magnetic field is described and is shown
to reduce the trigger rate to lower than the required level in Run~3.
A new trigger board, NSL, to integrate all the information from various detectors, has been developed,
and the implementation of the new logic on the FPGA has been successfully demonstrated.
\renewcommand{\thefootnote}{}
\footnote[0]{From ATL-DAQ-PROC-2018-007. Published with permission by CERN.}

\end{abstract}

\IEEEpeerreviewmaketitle

\section{Introduction}
\begin{figure}[b]
\centering
\includegraphics[viewport=0 0 602 270, clip, width = 3.2in]{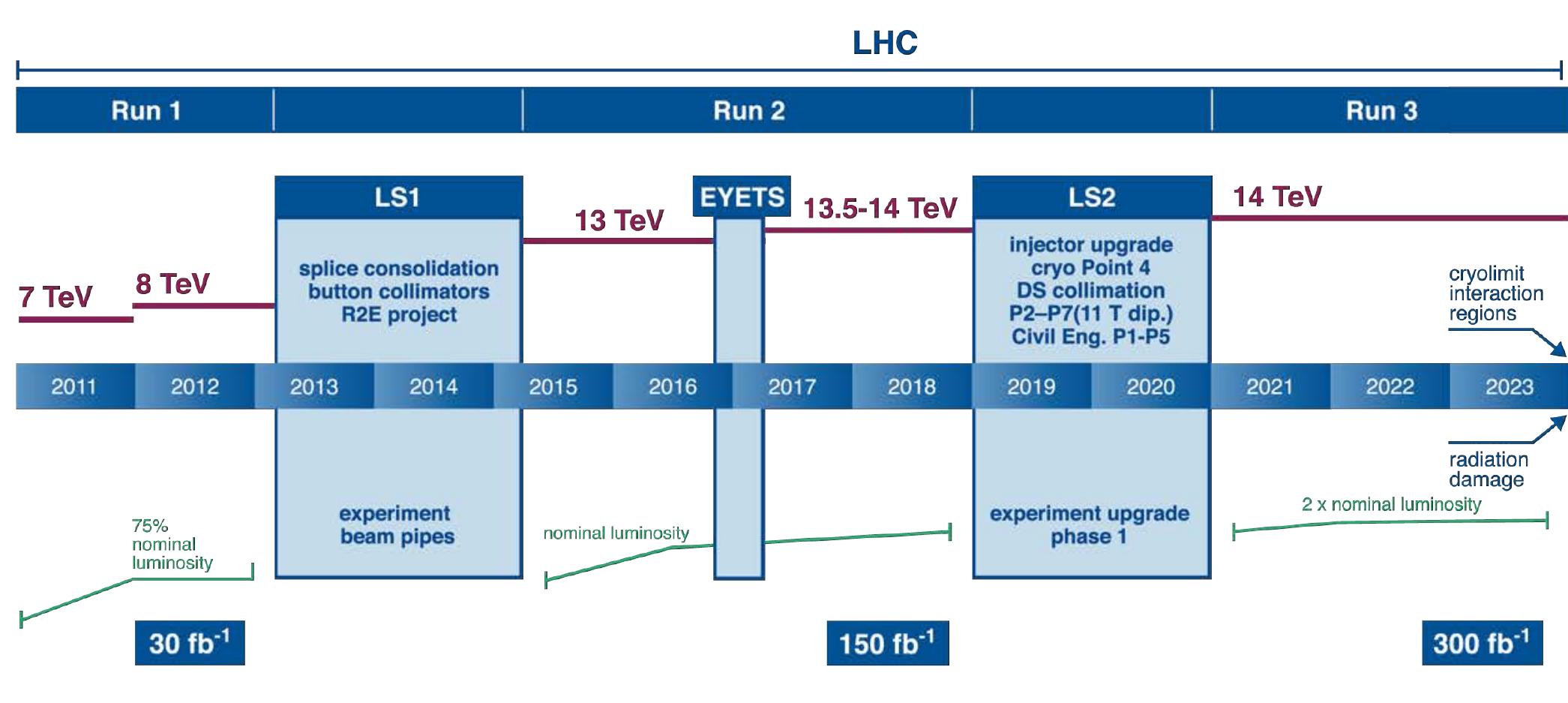}
\caption{LHC upgrade plans~\cite{ref:LHCschedule}. Run~3 will start from 2021, with $\sqrt{s}$ = 14~TeV and instantaneous luminosity of 3.0 \lumi}
\label{fig_LHC_plans}
\end{figure}

The Standard Model of particle physics offers the current best explanation of our universe, 
however there still remain some unsolved problems, such as the hierarchy problem and a lack of a dark matter candidate.
These mysteries imply the presence of a Beyond Standard Model (BSM) physics.
The collider experiment is one of the most promising approaches to probe BSM.
The Large Hadron Collider (LHC)~\cite{ref:LHC}, located in Switzerland, is a proton-proton collider with center-of-mass energy at $\sqrt{s} = 13$~TeV.
LHC provides the proton bunch collision at 40 MHz frequency, with its peak instantaneous luminosity at 2.1 \lumi.
LHC is planned to undergo a further upgrade of its luminosity and the center-of-mass energy, as shown in Fig.~\ref{fig_LHC_plans}.
Run~3 will start in 2021, with an instantaneous luminosity at 3.0 \lumi, 
which is about 1.5 times larger than the instantaneous luminosity in the current run (Run 2).

ATLAS~\cite{ref:ATLAS} is a multi-purpose detector to measure the particles produced at the interaction point (IP), 
located in one of the collision sites at LHC.
The final data recording rate of the ATLAS experiment is limited to 1~kHz, 
therefore an online event selection (trigger) is performed in order to decrease the event rate below the limitation,
while keeping the acceptance to new physics.
ATLAS trigger system consists of 2 parts; the Level-1 (L1) trigger and the High-Level trigger (HLT).
L1 trigger is a hardware-based trigger system that receives coarse detector information and reduces the event rate to smaller than 100~kHz.
For those events that passed the L1 trigger, the full-detector data is transferred to the Readout System,
and more precise information is passed to the HLT where final trigger decision is made.
The limitation of the L1 rate comes from the data transfer bandwidth, from the front-end electronics to the readout system.
This L1 rate limitation cannot be raised until a major update of the front-end electronics itself (which in fact will take place in Phase-2 Upgrade, from 2024).

Under this limitation, the requirements on the L1 trigger system will be severe at Run~3.
Despite the higher event rate, the L1 rate need to be kept at the same level.
The ATLAS detector needs an upgrade before LHC Run~3, 
to enhance its performance to cope with these high luminosity conditions.
Collectively this effort is known as the Phase-1 Upgrade.


\section{Phase-1 Upgrade of \\the ATLAS Level-1 Muon Trigger}
\begin{figure}[htbp]
\centering
\includegraphics[width = 3.2in]{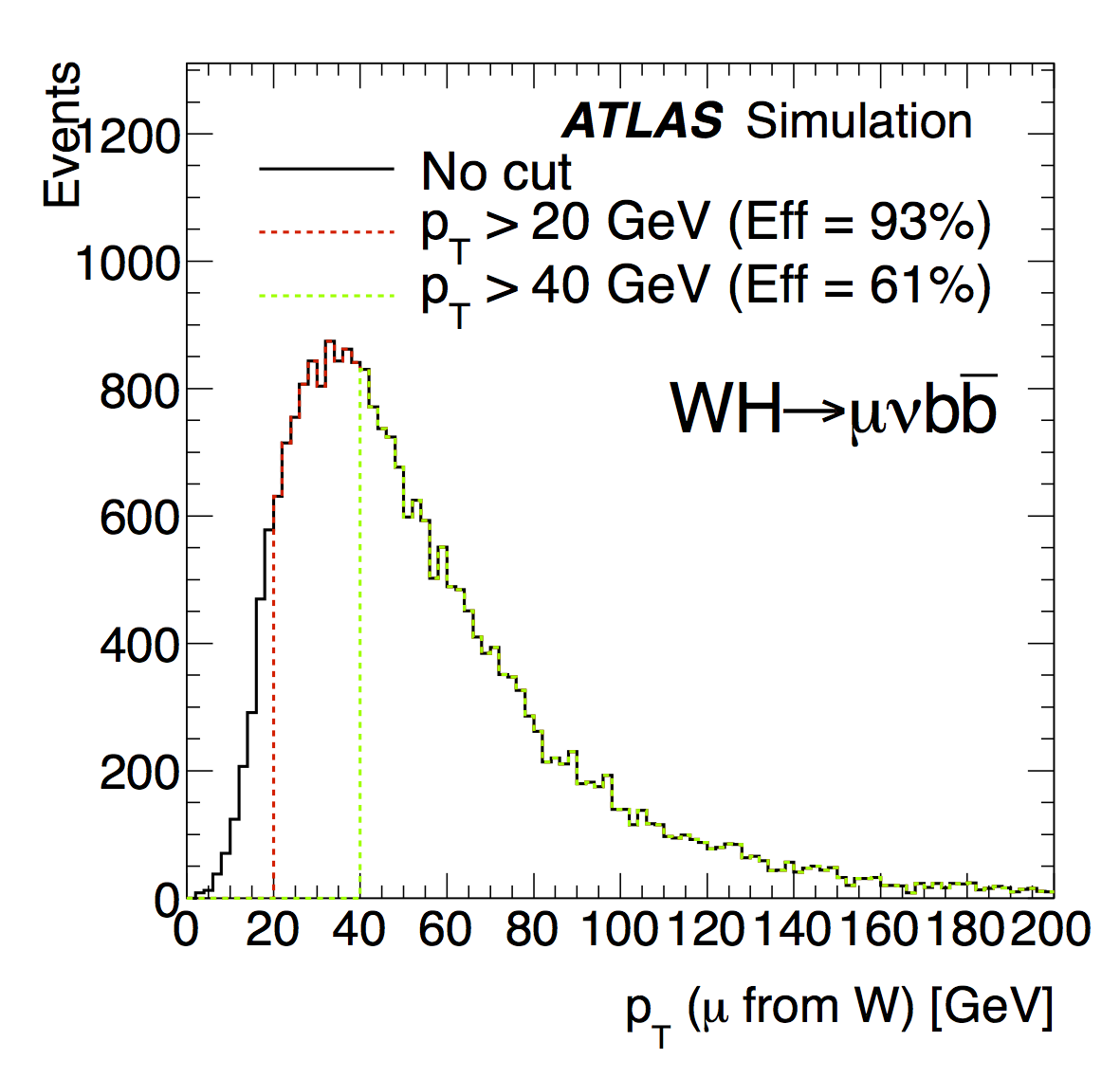}
\caption{\pT distribution of the muon from decay of a W boson produced in association with Higgs boson~\cite{ref:NSW_tdr}.
  93\% of the muons have \pT greater than 20~GeV, whereas only 61\% have \pT greater than 40~GeV.}
\label{fig_WH_W_lnu}
\end{figure}

The L1 trigger rate assigned to the un-prescaled lowest-threshold trigger for the single muon (primary muon trigger) in Run~3 has been defined to be 15~kHz, 
considering other triggers and physics requirements~\cite{ref:TDAQ_tdr}.
Current primary muon trigger has a transverse momentum (\pT) threshold of 20~GeV, with its trigger rate of 20~kHz at instantaneous luminosity of 2.0 \lumi.
With an extrapolation from the trigger rate at the current luminosity, the rate at Run~3 luminosity at 3.0 \lumi will be 30~kHz, which is twice as much as the requirement.
To keep the trigger rate below the requirements with current the trigger system, the \pT threshold would have to be raised to 40~GeV~\cite{ref:TDAQ_tdr}, 
which will immediately lead to decrease the physics acceptance significantly.
One of the examples that could be influenced strongly by the trigger threshold is the process where Higgs boson is produced in association with a W or Z boson.
In this production channel, for the Higgs decay modes $H \rightarrow b\bar{b}$~\cite{ref:Hbb} and $H \rightarrow c\bar{c}$~\cite{ref:Hcc}, 
the most important trigger will be the lepton trigger,
which triggers the events by identifying leptons from the W or Z decay.
If the muon trigger threshold is raised to 40~GeV, the trigger efficiency for $WH \rightarrow l\nu b\bar{b}$ will decrease by more than 30\%,
as shown in Fig.~\ref{fig_WH_W_lnu}.

In order to keep the primary muon trigger threshold to 20~GeV, upgrade of the L1 muon trigger is essential.
The main strategy of the upgrade is to reduce the muon trigger rate at regions with $|\eta| > 1.0$, the endcap region.

\begin{figure}[htbp]
\centering
\includegraphics[width = 3.2in]{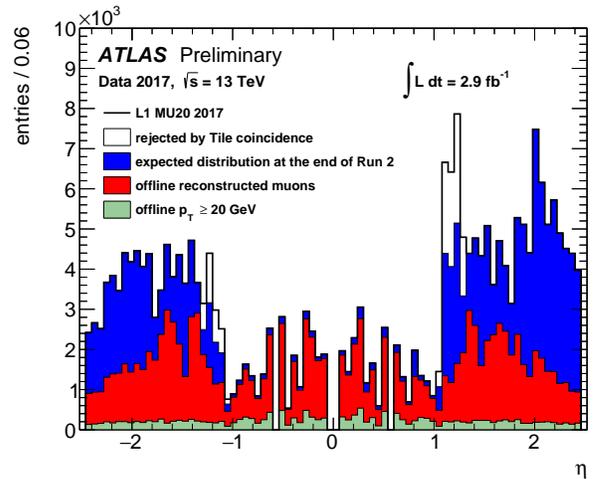}
\caption{Pseudo-rapidity ($\eta$) distribution of L1 MU20 candidate at Run~2~\cite{ref:pT_distribution}.
  The vertical axis corresponds to the trigger rate.
  Almost 80\% of the rate comes from the endcap region where $|\eta| > 1.0$.  
}
\label{fig_eta_Run2}
\end{figure}

Figure~\ref{fig_eta_Run2} shows the pseudo-rapidity ($\eta$) distribution of the Level-1 muon primary trigger candidate (L1 MU20).
This plot reveals that more than 90\% of the muon trigger candidates in the endcap region are due to background events,
i.e. events without muon with \pT above the threshold.
About 50\% of these background triggers are due to events with no associated reconstructed muon. 
These background triggers are known as ``fake'' triggers, caused by charged particles emerging from the beam pipe.
Other background triggers are due to muons with \pT below the threshold.
The main strategy of the upgrade is to eliminate the fake triggers and the low-\pT muons,
by implementing some new trigger algorithms which make use of several detectors that will be introduced in Run~3.

\begin{figure}[htbp]
\centering
\includegraphics[viewport=50 0 750 450, clip, width = 3.2in]{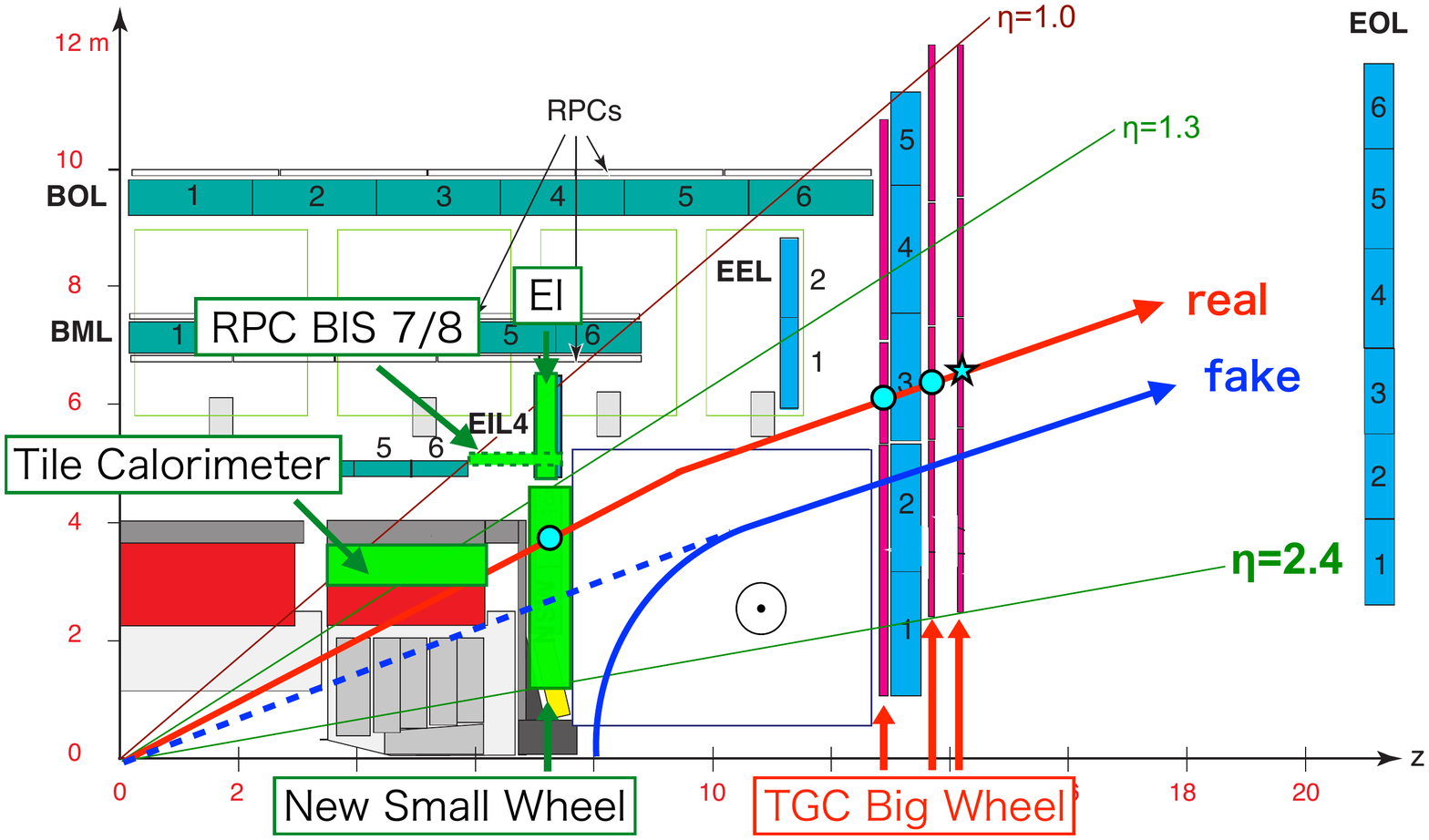}
\caption{Trigger strategy for Phase-1 Upgrade. Real muons from the IP leave hits in the detectors inside the magnetic field.
  On the other hand, fake particles emerged from the beam pipe do not leave hit in the detectors inside the field.}
\label{fig_phase1_strategy}
\end{figure}

The detectors for the endcap muon trigger in Run~3 are shown in Fig.~\ref{fig_phase1_strategy}.
The toroidal magnetic field bends the muon tracks, 
so that the \pT can be calculated from the track angles in the Thin-Gap Chambers (TGC)~\cite{ref:muon_spectrometer} installed outside the magnetic field.
These chambers are called TGC Big Wheel (TGC BW), as it forms a wheel-like structure.
As shown in Fig.~\ref{fig_phase1_strategy}, while the fake particles create muon-like tracks in the TGC BW,
they do not make any hits in the detectors inside the field.
Thus, the fake triggers can be rejected by requiring a coincidence between hits in the TGC BW and the detectors inside the field.
In Run~3, there are 4 detectors inside the  magnetic field that can be used in L1 muon trigger: 
New Small Wheel (NSW)~\cite{ref:NSW_tdr}, TGC EI~\cite{ref:muon_spectrometer}, RPC BIS 7/8~\cite{ref:BIS7/8}, and Tile Calorimeter~\cite{ref:Tile}.
NSW is a combined detector of micromegas~\cite{ref:MM} and sTGC (small-strip TGC), and will be newly installed in Run~3.
The position and angle resolutions in the information that NSW can provide to L1 system is 0.005 in the $\eta$ plane, 10 mrad in the $\phi$ plane, and 1 mrad in the $\theta$ plane.
These are dramatically high compared to the current inner chamber at the same position, which has resolutions of 0.15 in the $\eta$ plane and 65 mrad in the $\phi$ plane. 
TGC EI and Tile Calorimeter have already been installed since the beginning of the experiment, RPC BIS7/8 will be newly installed in Run~3.
The region with $1.3 < |\eta| < 1.9$ is covered by the NSW, and
the region with $1.0 < |\eta| < 1.3$ is covered by TGC EI, RPC BIS 7/8 and Tile Calorimeter each covering different $\phi$ regions.
Run~3 L1 endcap muon trigger will integrate information from all the detectors described above to make the final trigger decision.

\section{trigger algorithm}
\label{sec:Algorithm}
In this section, the algorithms of the Level-1 Endcap muon trigger in Run~3 is summarized.

\subsection{TGC BW local coincidence}
TGC BW consists of three stations, M1, M2 and M3, from the inner to the outer layers.
Hit segment in the outermost M3 station is used as the trigger seed.
If a line is drawn from the IP to a hit in M3,
this straight line corresponds to infinite-momentum track.
The deviations from this path in the M2 and M1 planes are used to calculate \pT.
For the lower \pT muons, the tracks will be bent more by the magnetic field,
so the deviations from the straight track become larger.
This trigger logic is implemented on a FPGA chip by a pre-defined Look-Up-Table.
The deviation in the $R$ direction, d$R$, and the deviation in the $\phi$ direction, $\rm{d}\phi$, 
are handed to the LUT, and then the LUT immediately returns the \pT value.
The correlation between the deviation angle and the \pT is quite different depending on the trigger seed position,
because the toroidal magnetic field is not uniform. 
Therefore the LUT must be defined depending on the position, to maximize the performance.

\subsection{Position Matching}
\begin{figure}[htbp]
\centering
\includegraphics[viewport=120 0 800 500, clip, width = 3.2in]{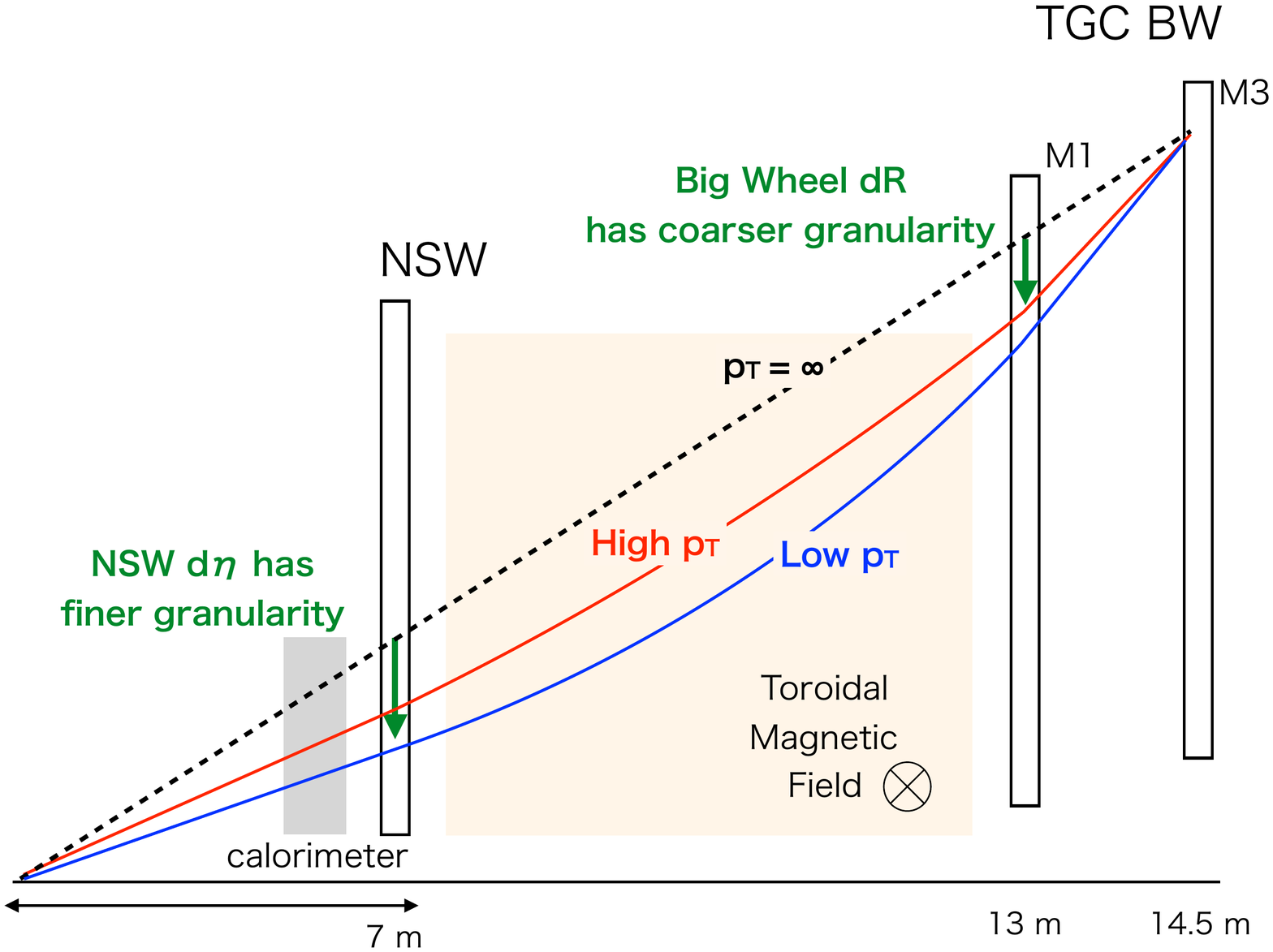}
\caption{Schematic diagram for the position matching algorithm. 
  With a good position resolution at inner station, the segment position at inner station will have better sensitivity to the muon \pT.}
\label{fig_NSW_pos}
\end{figure}

The main concept of the position matching with the inner detector is simple: 
that is, when the TGC BW finds a trigger candidate by its local coincidence, 
to confirm the decision by requiring hit in detectors inside the magnetic field.
The position matching algorithm requires hits at appropriate position in the inner detector.
As shown in Fig.~\ref{fig_NSW_pos}, position matching will not only reject the fake triggers, 
but also reject low-\pT candidates if the resolution of the inner detector is high enough.
NSW, with the largest coverage in endcap region and high granularity, 
is expected to have an impact on the trigger rate by rejecting low~\pT muons.

\subsection{Angle Matching}
\begin{figure}[htbp]
\centering
\includegraphics[viewport=120 0 800 500, clip, width = 3.2in]{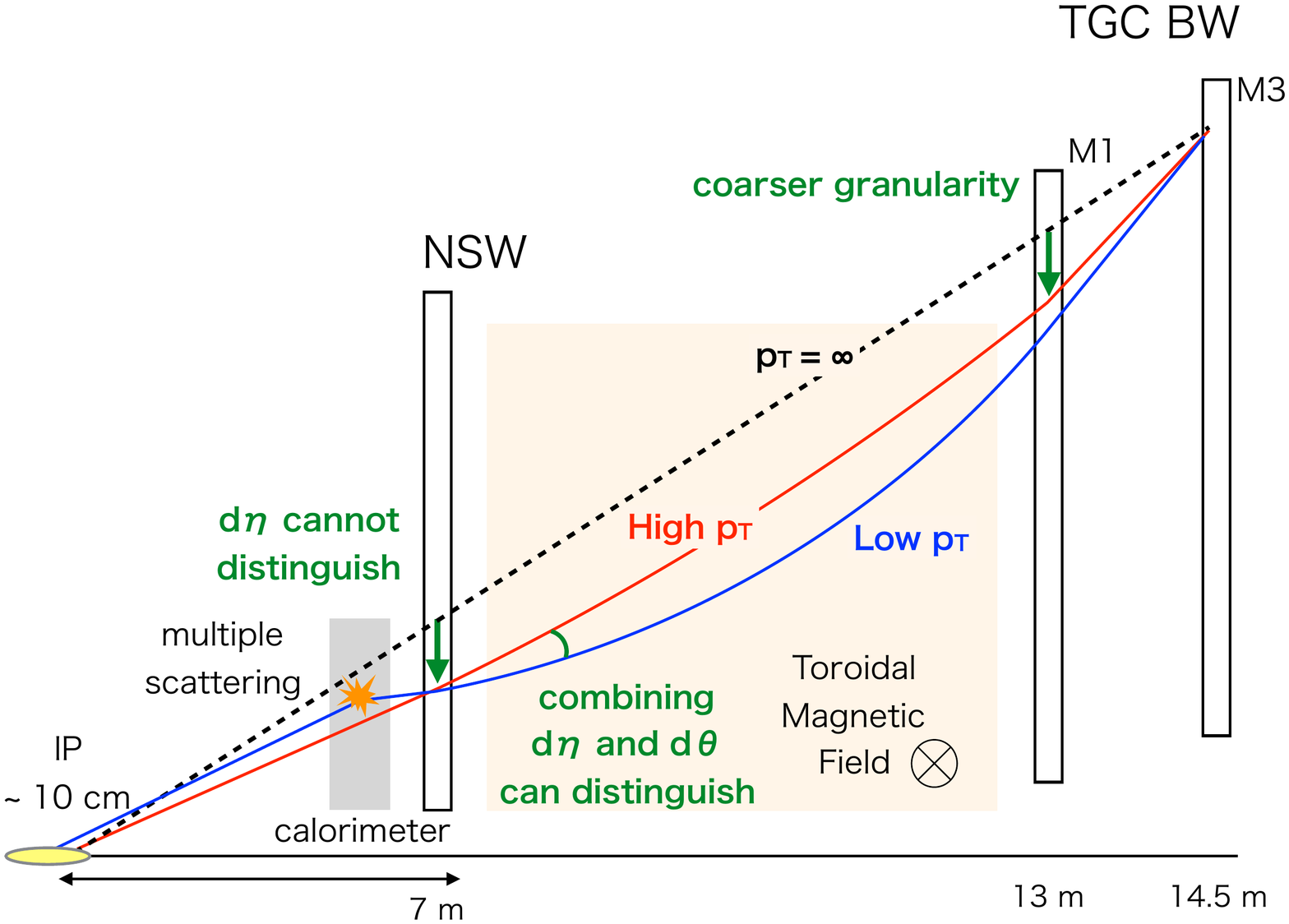}
\caption{Schematic diagram for the angle matching algorithm. 
  By combining the position and angle information at inner station, it is possible to further distinguish low-\pT muons with straight-like hit position.
  }
\label{fig_NSW_angle}
\end{figure}

In addition to the position matching algorithm, it is also possible to make use of the angle information at the inner station (angle matching).
In the NSW trigger processor, d$\theta$ is defined as the angle of the segment with respect to the direction pointing at the detector center.
For a muon emitted at the detector center and arrived to the inner station with a straight track, d$\theta$ should be ideally zero.
However, we need to consider the beam spot size in $z$-direction (O(10cm)) and the multiple scattering with the detector materials, 
especially in the calorimeters.
These effects allow the low-\pT muons to fake the high-\pT muons, as explained in Fig.~\ref{fig_NSW_angle}.
Here, a low-\pT muon is indicated by the blue line, and high-\pT muon is indicated by the red line.
In Fig.~\ref{fig_NSW_angle}, we are considering the situation when there is a trigger seed in a specific position at M3 plane.
The low-\pT track can fake the hit position of the high-\pT tracks if it was scattered in the calorimeter region.
The hit position in NSW and BW are very similar, but in this case the d$\theta$ information should differ significantly.
Note that d$\theta$ information on itself cannot distinguish the \pT of the muon, whereas can be combined with hit position in NSW and BW
to retrieve more accurate \pT information.
By combining d$\theta$ and the position information, especially the $\eta$ position at NSW, low-\pT muons can be eliminated effectively.

\subsection{Trigger performance}
\begin{figure}[htbp]
\centering
\includegraphics[width = 3.2in]{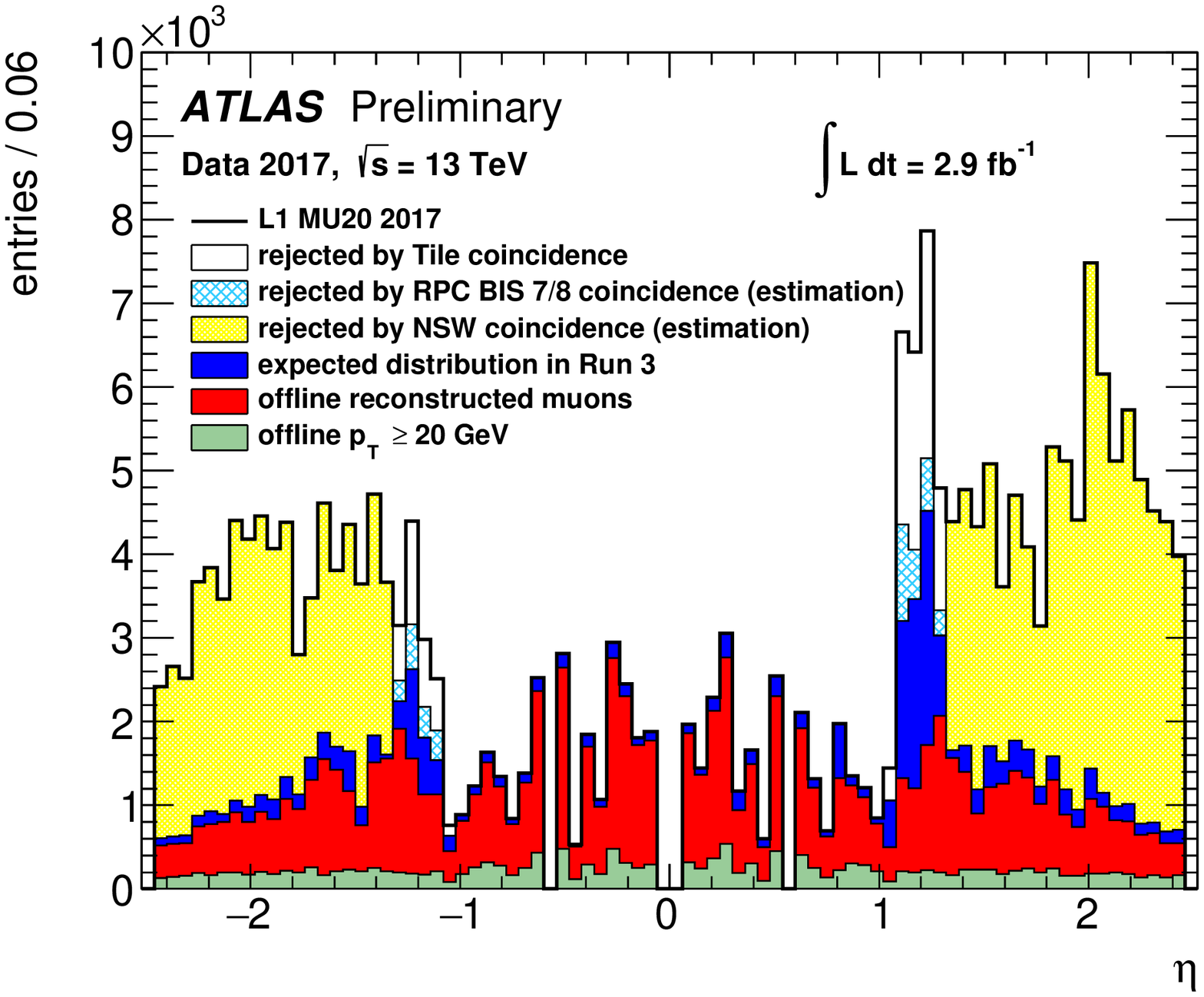}
\caption{Trigger efficiency relative to Run 2 trigger, as a function of \pT~\cite{ref:pT_distribution}.
  The NSW segment reconstruction efficiency is assumed to be 97\%.
  ``Run-2 (BW + FI) BW + NSW(d$\eta$:d$\phi$)'' corresponds to applying position matching algorithm,
  ``Run-2 (BW + FI) BW + NSW(d$\eta$:d$\phi$ \& d$\eta$:d$\theta$)'' corresponds to applying both position and angle matching algorithms.}
\label{fig_turnon}
\end{figure}

The trigger efficiency as a function of \pT of the offline-reconstructed muon is shown in Fig.~\ref{fig_turnon}.
This plot shows the relative trigger efficiency compared to the Run 2 trigger efficiency,
calculated using single muon MC samples.
The NSW track segment reconstruction efficiency is assumed to be 97\%, and is included in this study.
A significant increase in trigger rejection power for low-$p_{\rm T}$ muon candidates are seen, 
for example additional 50\% of the 10~GeV muons are rejected by taking both position and angle matching.

\begin{figure}[htbp]
\centering
\includegraphics[width = 3.2in]{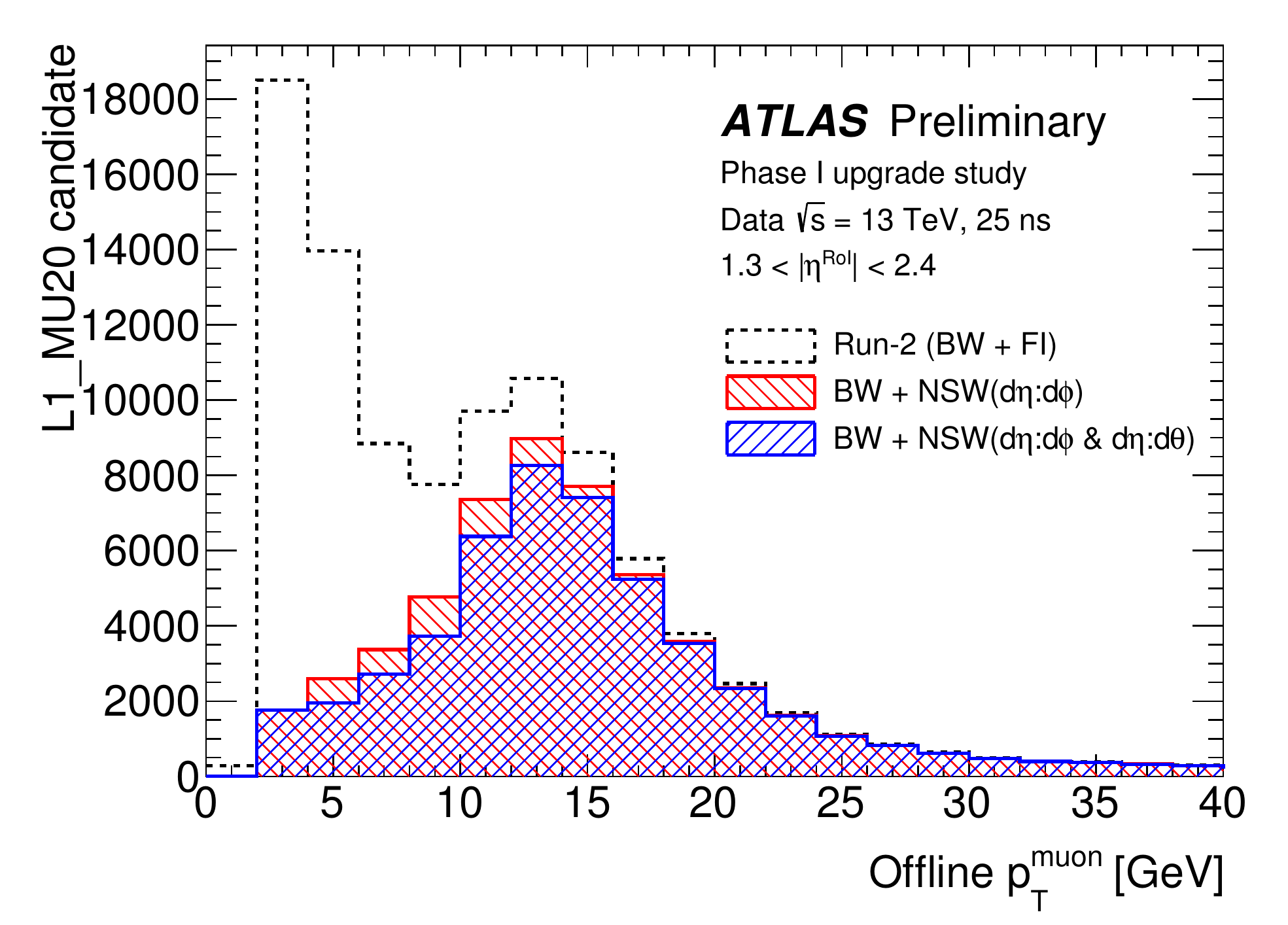}
\caption{\pT distribution of the muons that passed Level-1 muon primary trigger~\cite{ref:pT_distribution}.
  The distribution in dashed line is obtained from 2016 data.
  The distributions after applying position and angle matching are calculated by multiplying relative efficiency estimated by MC.}
\label{fig_pt_distribution}
\end{figure}

Estimation of the trigger rate is shown in Fig.~\ref{fig_pt_distribution}.
The original \pT distribution of the muons that pass the L1 MU20, shown by the dashed black line, is retrieved from the 2016 data.
The detailed conditions can be found in \cite{ref:pT_distribution}.
Note that the fake triggers are not included in this plot, as the fake triggers are defined as triggers that do not match with offline reconstructed muons,
and so that the \pT of the muons cannot be defined.
The distribution after the NSW coincidence is obtained by multiplying the relative trigger efficiency 
in Fig.~\ref{fig_turnon} to the original distribution.

\begin{figure}[htbp]
\centering
\includegraphics[width = 3.2in]{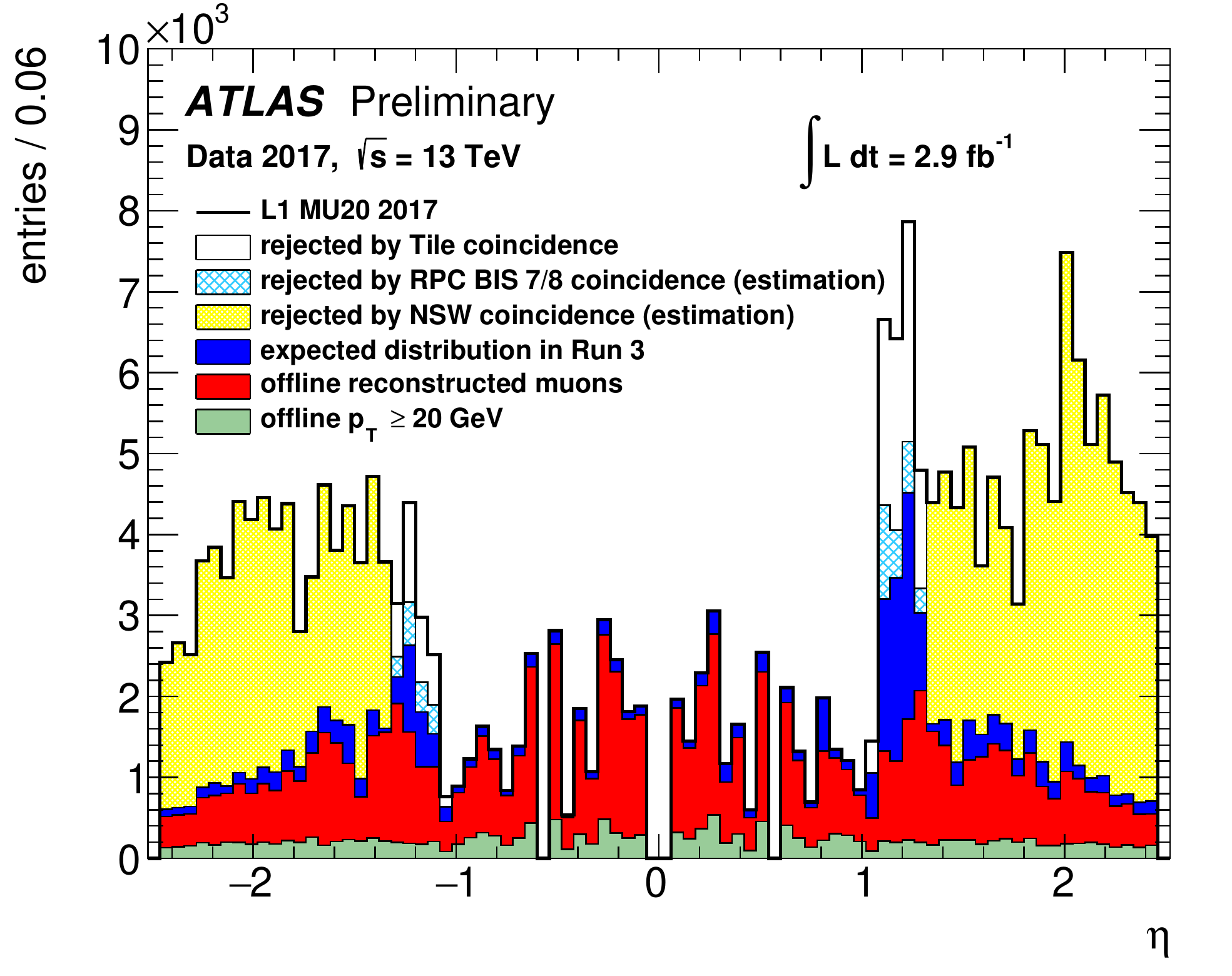}
\caption{Pseudo-rapidity ($\eta$) distribution of L1 MU20 candidate~\cite{ref:pT_distribution}.
  The fake reduction by NSW and BIS 7/9 is estimated using other muon detector segments, while low-\pT rejection is estimated from MC.}
\label{fig_eta_Run3}
\end{figure}

Figure~\ref{fig_eta_Run3} shows the eta distribution of the L1 MU20 trigger seed position.
The fake rejection power of NSW and RPC BIS7/8 coincidence was estimated by using 
the muons segments in the other detectors.
The low-\pT trigger rejection is estimated from the MC study above.
The EI coincidence is already included in Run 2, 
and the Tile Calorimeter coincidence performance is estimated from 2017 collision data.
The trigger rate, with all the coincidence logics included,
is estimated to be 14.2~kHz at 3.0 \lumi, which meets the Run~3 requirements.

\section{Development of the New Sector Logic board}
\begin{figure}[tbp]
\centering
\includegraphics[viewport=0 0 280 200, clip, width = 3.4in]{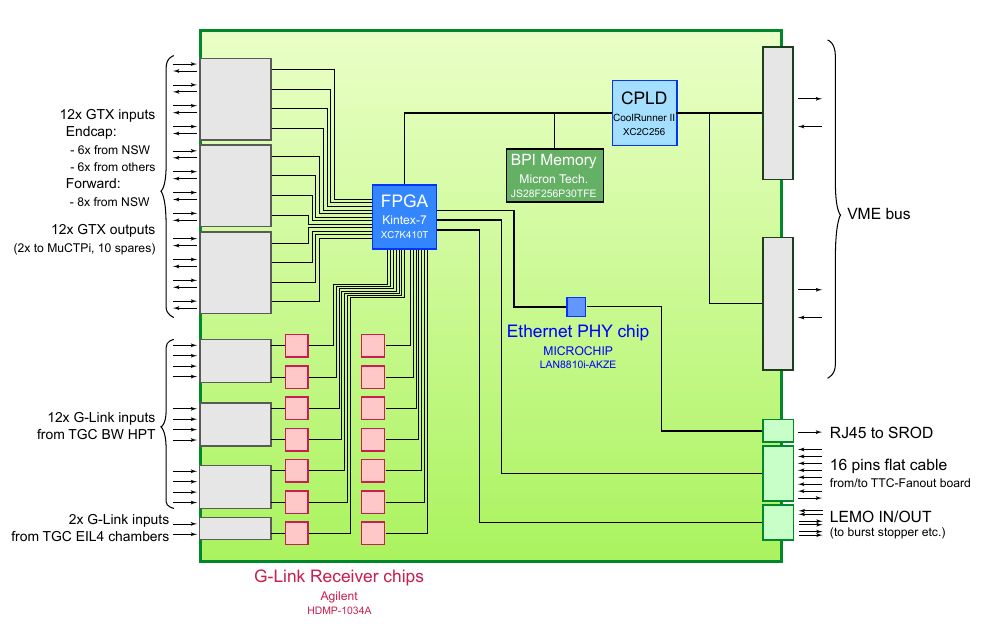}
\caption{New Sector Logic (NSL) board block diagram.
  NSL is a VME 9U board, with Xilinx Kintex-7 FPGA for the main processor.}
\label{fig_NewSL_block_diagram}
\end{figure}

Figure~\ref{fig_NewSL_block_diagram} shows the block diagram of the new Sector Logic board (NSL).
In this section, we summarize the hardware specification of the NSL board.

\paragraph{FPGA}
The main processor FPGA of NSL is required to have many user I/Os, mainly for the 14 G-LINK~\cite{ref:GLINK} connections.
These G-LINK connections are used to receive data from TGC BW.
G-LINK uses 21 input user I/Os per channel, including the control/flag bits, 
therefore 14 G-Links will use 294 input I/O ports.
Adding other I/O ports, the total user I/O required for New Sector Logic FPGA will be greater than 400. 
Another requirement for the FPGA is to have some Multi-Gigabit transceivers
in order to receive data from NSW and other detectors.

Kintex-7 FPGA~\cite{ref:7Series} XC7K410T-1FFG900 made by Xilinx Inc. was chosen as a main processor.
XC7K410T has 500 user I/O pins, which is the largest number in the Kintex-7 series.
XC7K410T supports GTX~\cite{ref:GTX}, and contains 16 GTX transceivers.
GTX is a multi-gigabit transceiver implemented on Xilinx Kintex-7 series FPGA.
16 transceiver port is enough to fulfill the requirements.

This FPGA has 795 Block RAMs (BRAMs)~\cite{ref:BRAM}, which is approximately 20 times as many as the current Sector Logic.
BRAM is a large memory block which can contain up to 38 Kb of data, and is used to implement trigger LUTs in Sector Logic.
Large number of BRAMs means a larger capability of trigger logic implementation.
The BRAM resource needed to implement the trigger logic will strongly depend on what kind of logic we will use, 
and also on how we implement them on the BRAMs.

\paragraph{CPLD}
CPLD is used for the VME bus control.
Xilinx CoolRunner-II XC2C256-7Q208C~\cite{ref:CPLD} has been chosen, 
which is the same chip as the one used for the current Sector Logic.
Because the CPLD is a non-volatile memory, FPGA on the NSL board can be 
configured via VME as soon as the power is turned on.
FPGA configuration via BPI is also controlled by CPLD.

\paragraph{BPI}
Micron JS28F256P30T has been chosen for memory to contain a firmware design.
This memory can contain up to 256 Mb data, which is enough to contain the firmware design.

\paragraph{G-LINK Receiver chip}
HDMP-1034A chip~\cite{ref:GLINK} has been selected for G-LINK receiver chip.
This chip receives serial data in G-LINK protocol, deserialize then transmit it as parallel signals.
Same chip as the one used in current Sector Logic has been chosen.

\paragraph{Ethernet PHY chip}
Microchip Tech. LAN8810i-AKZE~\cite{ref:PHY} is used for ethernet connection.
Using SiTCP~\cite{ref:SiTCP} technology to connect this PHY chip to FPGA appropriately will 
allow data transmission by TCP/IP.

\paragraph{Clock Jitter Cleaner}
Silicon Labs Clock Jitter Cleaner Si5334~\cite{ref:Si5334} is used to create a reference clock for the GTX.
The chip is semi-customized (Si5334C-B05812-GM) to input frequency of 40.08 MHz, which is the exact LHC bunch crossing frequency.
The output of this jitter cleaner is a low-jitter 160 MHz clock, which is used for the GTX reference clock.

\section{Firmware Implementation and Performance}
The trigger logic explained in Section~\ref{sec:Algorithm} need to be processed on the FPGA, to be used in the L1 trigger.
The requirement on the logic is that the trigger decision must be processed in fixed and short latency.
The latency from the collision to the L1 trigger output of the Sector Logic is limited to 53 LHC clocks, including all the latency such as 
TOF, digitization, signal serialization, data transfer, trigger logic etc.
The latency that can be assigned to inner coincidence logic is, considering other latencies, limited to 2 LHC clocks.

On the FPGA, we need to implement 3 types of LUTs: TGC BW local coincidence, position coincidence between BW and inner stations,
and angle matching between BW and inner stations.
A problem is that there can be several inner coincidence candidates for one BW trigger seed.
For example, for the NSW segments, 16 candidates at the maximum can come into the scope for one BW trigger seed.
To implement this kind of logic in a fixed-latency scheme, 
a naive method would be to implement 16 same LUTs so that the trigger processing can be performed in parallel.
In such an implementation, we would need 16 times the FPGA resource, which is not acceptable in the current chip.
Another way is to process the 16 candidates in serial, one-by-one.
In this way the resource usage can be minimized, however the latency will be 16 times longer than the original latency.

\begin{figure}[htbp]
\centering
\includegraphics[viewport=50 0 800 350, clip,width = 3.2in]{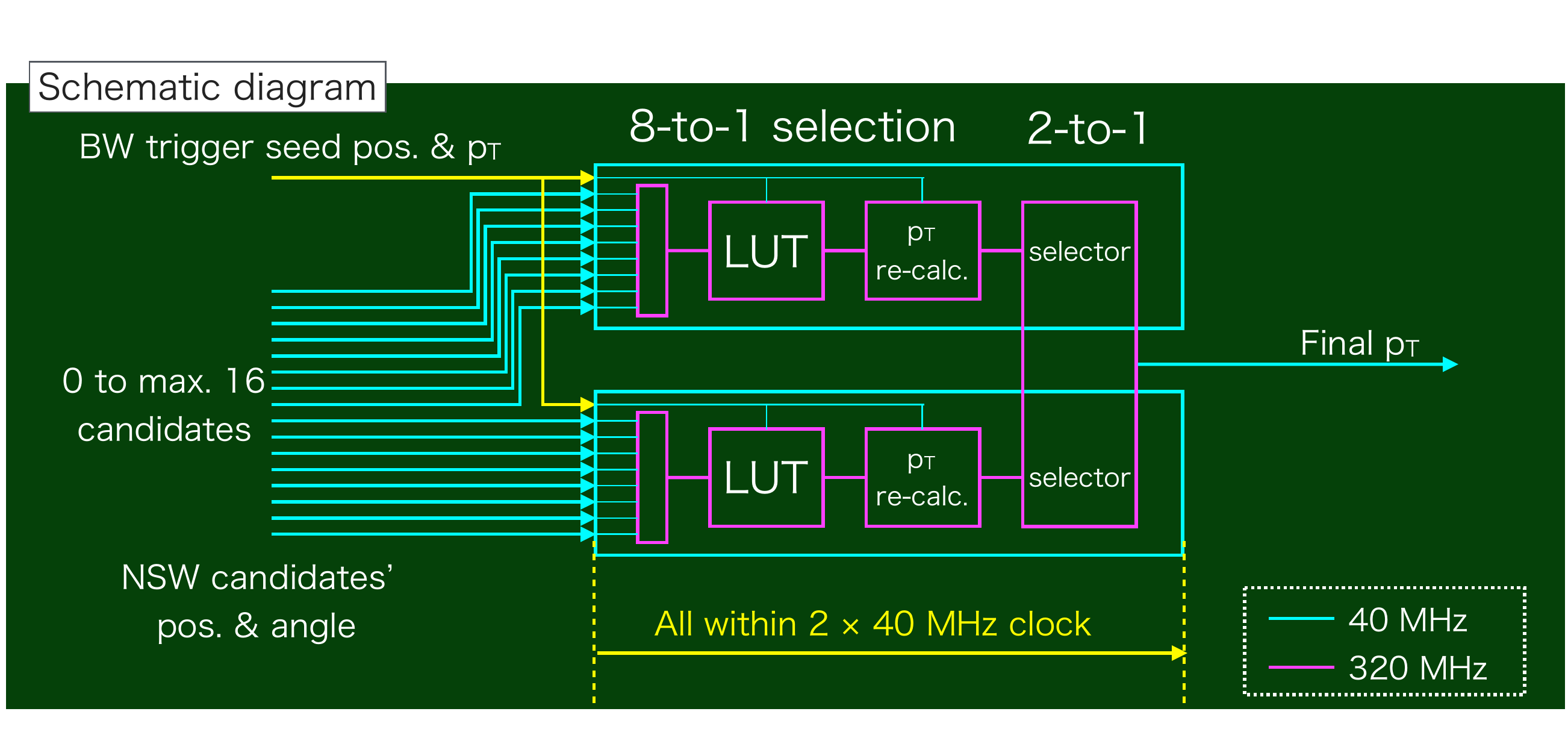}
\caption{LUT implementation schematic.
  The different colors of the lines in the figure shows different clock domains.
  Up to 16 NSW candidates can be provided for one BW trigger seed, 
  and they are split into 2 groups and are handed to two LUTs which have exactly the same content.
  The LUT is operated at 320 MHz clock to process all the candidates in short latency.
  The output of the LUT is then compared in two steps, and the highest \pT candidate will be chosen as the final output.
}
\label{fig_LUT_schematic}
\end{figure}

To overcome these problems, we have decided to pick the good points of the above two ideas.
Fig.~\ref{fig_LUT_schematic} shows the brief schematic diagram of the implementation.
In our implementation, two identical LUTs (LUT pair) were placed in parallel.
Each LUT is operated at 320 MHz clock, which is 8 times faster compared to the LHC clock.
Therefore, the LUT can process 8 candidates in one LHC clock.
The 2 $\times$ 8 output values are then compared, and the highest \pT candidate is chosen as the final output.
This selection is done in 2 steps: in Fig.~\ref{fig_LUT_schematic} they are referred to as ``\pT re-calc.'' and ``selector''.
The first step simply compares the output of the 8 candidates from one of the LUTs, and then choose a highest \pT candidate.
The second step, ``selector'', compares the 2 candidates from each LUT ``\pT re-calc.'' algorithm, 
and then send the higher one back on the 40 MHz clock domain, as the final \pT value.
This implementation enables us to process maximum 16 candidates within 2 LHC clocks, and also within reasonable amount of RAM resource usage.

\begin{figure}[htbp]
\centering
\includegraphics[width = 3.2in]{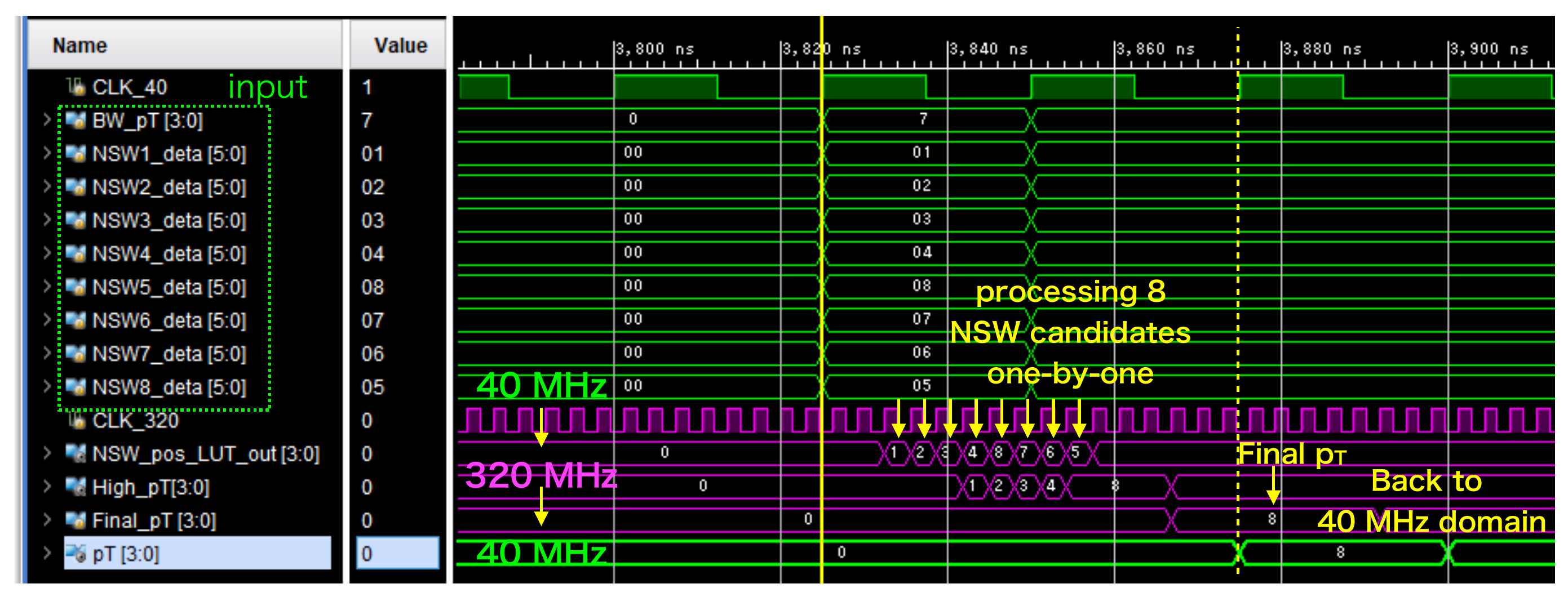}
\caption{An example of the LUT simulation results.
  The different colors of the lines in the figure shows different clock domains.
  ``BW\_\pT'', is the \pT calculated by TGC, and ``NSW1\_deta'' to ``NSW8\_deta'', 
  correspond to d$\eta$ values of the eight NSW segments.
  These information are handed to the logic as input data.
  The LUT output is referred to as ``NSW\_pos\_LUT\_out'', which, in this simulation, only returns the input variable.
  The selected final \pT is given back to the 40 MHz clock domain, after 2 LHC clocks.
}
\label{fig_LUT_simulation}
\end{figure}

Fig.~\ref{fig_LUT_simulation} shows a Vivado~\cite{ref:Vivado} simulation output of the trigger logic part.
This test was performed under a very simple situation, where only one of the LUT pair is used.
The LUT is initialized to simply return the input value.
The different to colors of the lines in the figure shows different clock domains.
In this simulation, the original \pT calculated in the BW local coincidence logic is ``7'', as shown in ``BW\_pT'' column.
Eight inner segments information are handed to the LUT, shown as ``NSW1\_deta'' to ``NSW8\_deta''.
The 8 candidates are handed to the 320 MHz clock domain in series, and are processed one-by-one, from ``NSW1\_deta'' to ``NSW8\_deta'', by the same LUT.
The output of the LUT is referred to as ``NSW\_pos\_LUT\_out'' in Fig.~\ref{fig_LUT_simulation}.
These output values are then compared, and the maximum \pT value is given to the ``High\_pT'' register.
The ``High\_pT'' value is chosen as ``Final\_pT'', and then are handed back on the LHC 40 MHz clock domain.
All these processes are completed within two LHC clocks.

\enlargethispage{-1.3in}
We have performed similar tests for various input patterns and LUT initialization values.
These tests were done on logic simulation by Vivado, and also on an actual FPGA chip.
The output of the trigger logic was consistent with our expectations, 
thus it is confirmed that we succeeded in handling the coincidence logic for maximum 16 inner segments, 
in a fixed latency and reasonable FPGA resource consumption.

\section{Conclusion}
Phase-1 upgrade of the ATLAS Level-1 endcap muon trigger is essential to keep the physics acceptance at LHC Run~3.
A new trigger algorithm to take coincidence with detectors inside- and outside- the magnetic field was suggested.
The new algorithm consists of position and angle matching, to reduce the trigger rate to 14.2~kHz at 3.0 \lumi,
which meets the Run~3 requirements.
A new trigger board, NSL, to integrate all the information from various detectors, was developed.
NSL has 12 GTX input port and 14 G-LINK input port, to receive data from all the five detectors; 
TGC BW, NSW, TGC EI, RPC BIS7/8 and Tile Calorimeter.
The firmware implementation of the position and angle matching is realized by combining two ideas.
The first is to place identical LUTs in parallel, which allows to process two candidates at the same time.
The second is to re-use the LUT 8 times each, on a fast 320~MHZ clock, so that 16 candidates can be processed within 
the limited latency of two LHC clocks, while keeping the resource usage to acceptable level.
The test of the trigger logic was successfully done on simulation, as well as on the real NSL board.






\bibliographystyle{IEEEtran}
\bibliography{IEEEabrv,./bibliography.bib}

%


\end{document}